\documentclass[prl, aps, twocolumn, groupedaddress, superscriptaddress]{revtex4}

\usepackage{slashed}
\usepackage{graphicx}
\usepackage{subfigure}
\usepackage[usenames, dvipsnames]{color}
\usepackage{graphics}
\usepackage{hyperref}
\usepackage{bm}
\usepackage{amsfonts}

\hypersetup{backref,
colorlinks=true,
linkcolor=blue,
linktoc=page,
citecolor=blue}

\begin{document}

\title{Non-Abelian Gauge Fields in Photonic Cavities and Photonic Superfluids}

\author{H. Ter\c{c}as}
\affiliation{Institut Pascal, PHOTON-N2, Clermont Universit\'e, Blaise Pascal University, CNRS,24 Avenue des Landais, 63177 Aubi\`ere Cedex, France}
\author{H. Flayac}
\affiliation{Institut Pascal, PHOTON-N2, Clermont Universit\'e, Blaise Pascal University, CNRS,24 Avenue des Landais, 63177 Aubi\`ere Cedex, France}
\affiliation{Institute of Theoretical Physics, \'{E}cole Polytechnique F\'{e}d\'{e}rale de Lausanne EPFL, CH-1015 Lausanne, Switzerland}
\author{D. D. Solnyshkov}
\affiliation{Institut Pascal, PHOTON-N2, Clermont Universit\'e, Blaise Pascal University, CNRS,24 Avenue des Landais, 63177 Aubi\`ere Cedex, France}
\author{G. Malpuech}
\affiliation{Institut Pascal, PHOTON-N2, Clermont Universit\'e, Blaise Pascal University, CNRS,24 Avenue des Landais, 63177 Aubi\`ere Cedex, France}

\begin{abstract}
We show that the interplay between the structure anisotropy and the energy splitting between the TE and TM modes of a microcavity leads to the appearance of a Non-Abelian gauge field for a propagating polariton condensate. The field texture can be tuned by rotating the sample and ranges continuously from a Rashba to a monopolar field. In the linear regime, the monopolar field leads to a remarkable focusing and conical diffraction effects. In the interacting regime, the spin-orbit coupling induces a breakdown of superfluidity. The spatially homogeneous flows become unstable and dynamically evolve into textured states such as stripes and domain walls.
\end{abstract}

\maketitle

The creation of synthetic gauge fields has been seriously addressed in the last years \cite{lin1, lin2, aidelsburger, chen}. In condensed matter systems, gauge fields due to spin-orbit coupling (SOC) play a central role in the spin-Hall effect \cite{hall1, hall2}, topological insulators \cite{ti1,ti2} and semiconductor-based spintronics \cite{koralek}. Artificial magnetic fields in atomic Bose-Einstein condensates (BECs) allow the formation of vortices without stirring \cite{lin1}, while an artificial field of the Rashba type leads to the formation of spin domains \cite{lin2}. In purely photonic systems synthetic gauge fields have been proposed \cite{Carusotto,Hafezi} and produced \cite{Hafeziarxi},  being in the basis of topologically protected states of a spin Hall system. Fang et al. \cite{Fang} have shown that the photonic de Haas-van Alphen effect is possible in the presence of synthetic magnetic fields \cite{Rechtman}. Photonic topological insulators have been experimentally demonstrated \cite{Rechtmannature,Jia}.

\par In optical cavities, and generally in confined photonic structures, both photons and mixed particles such as exciton-polaritons experience spin-orbit coupling. It originates from the splitting between TE and TM modes \cite{review, kuther} and is characterized by a $k^2$ scaling and a $2\theta_k$ dependence, where $\theta_k$ is the angle between the wave-vector $\mathbf{k}$ and the horizontal. An additional splitting between linear polarizations at $k=0$ corresponds to a constant effective magnetic field associated with the crystallographic axes \cite{Klopo}. While the TE-TM field leads to several fascinating effects, such as the optical spin-Hall effect \cite{OSHE, Leyder, hugof} or the acceleration of magnetic monopole analogs \cite{Soln, Hivet}, it cannot be integrated in the Hamiltonian as a minimal coupling. In other words, alone, it does not exhibit the properties of a true gauge field satisfying gauge transformations. The dispersion of the elementary excitations of a polariton BEC in the presence of the TE-TM splitting is anisotropic \cite{shelykh}, but the dynamics of the condensate is completely stable. The physical reason is that the TE-TM field does not break the spatial inversion symmetry, contrary to the Rashba/Dresselhaus field \cite{dalibard,rashba, dresselhaus}. In atomic condensates with linear in $k$ SOC, the ground state in the homogeneous limit is either plane-waves or stripes \cite{wang}. When a trapping potential is present, the two phases still exist as ground state in some limiting cases, in addition to the vortex phase which is specific of trapped system  \cite{santos, hu}. 
In contrast, the TE-TM field in polariton BECs allows a homogeneous ground state.

\begin{figure}[ht!]
\centering
\includegraphics[scale=0.20]{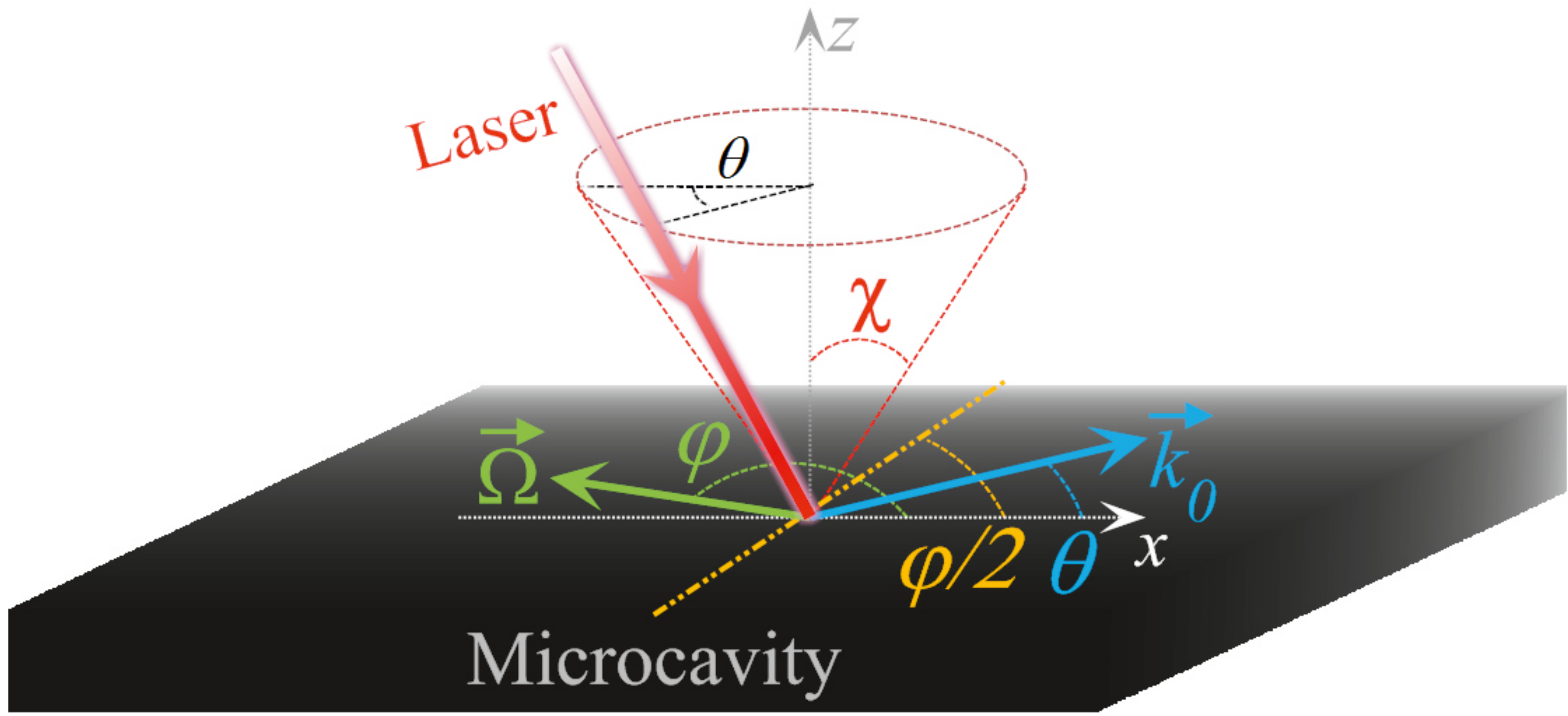}
\caption{(color online). Scheme of the experimental configuration necessary to produce a controllable gauge field in a microcavity. The angle between the crystallographic axis (the effective magnetic field $\bm \Omega$) and the horizontal is $\varphi/2$ ($\varphi$). A laser with a vertical inclination $\chi$ making an angle $\theta$ with the horizontal produces a propagating polariton beam with wave-vector $\mathbf{k}_0=(\omega_0/c)\sin\chi(\cos\theta,\sin\theta)$, where $\omega_0$ is the laser frequency and $c$ is the light speed.}
\label{fig1}
\end{figure}

In this Letter, we show how to create an effective gauge field for propagating photons or polariton condensates combining the TE-TM splitting and the in-plane field of the microcavity. At some critical velocity, the dynamics of the low-wavevector excitations is governed by a non-Abelian gauge field of the Rashba kind. In the non-interacting photonic regime, a spin splitting scaling linearly in $k$ leads to focusing and conical diffraction effects \cite{Hamilton, Lloyd,Berry,Peleg}. In the interacting regime, the same field makes the homogeneous superfluid flow unstable, in close analogy with the atomic condensate instability in the presence of synthetic gauge fields. A flowing condensate can be created at a controlled wavevector $\mathbf{k}_0$ by simply adjusting the angles between the laser beam and the normal and horizontal axes of the cavity (see Fig. \ref{fig1}). Around a critical velocity $\mathbf{v}_*=\hbar \mathbf{k}_*/m$, for which the static $\bm \Omega$ and TE-TM fields exactly compensate, the effective field acting on the pseudo-spin takes the shape of Rashba or monopolar fields. In the present discussion, $\bm \Omega= \Omega (\cos\varphi,\sin\varphi)$ describes an effective in-plane magnetic field associated with the crystallographic axis \cite{review}, as illustrated in Fig. \ref{fig1}. The field orientation in the $(k_x,k_y)$ plane can be tuned by simply rotating the sample. By investigating the properties of the collective excitations of a condensate polarized along the static field, we show that the gauge field makes the homogeneous superfluid flow unstable i) against transverse excitations in a limited region below $k_*$ and ii) against both transverse and longitudinal excitations for wave vectors above $k_*$, depending on the condensate polarization. These two instability mechanisms trigger the formation of spin stripes and  domain walls. Remarkable features occur when the polarization of the condensate is perpendicular to the static field. In such a case, a slowly propagating flow is unstable, whereas the superfluid motion is recovered above a critical velocity. We find the latter case to be an unexpected breakdown of superfluidity in the subsonic regime. \par
\begin{figure}[t!]
\centering
\includegraphics[scale=0.55]{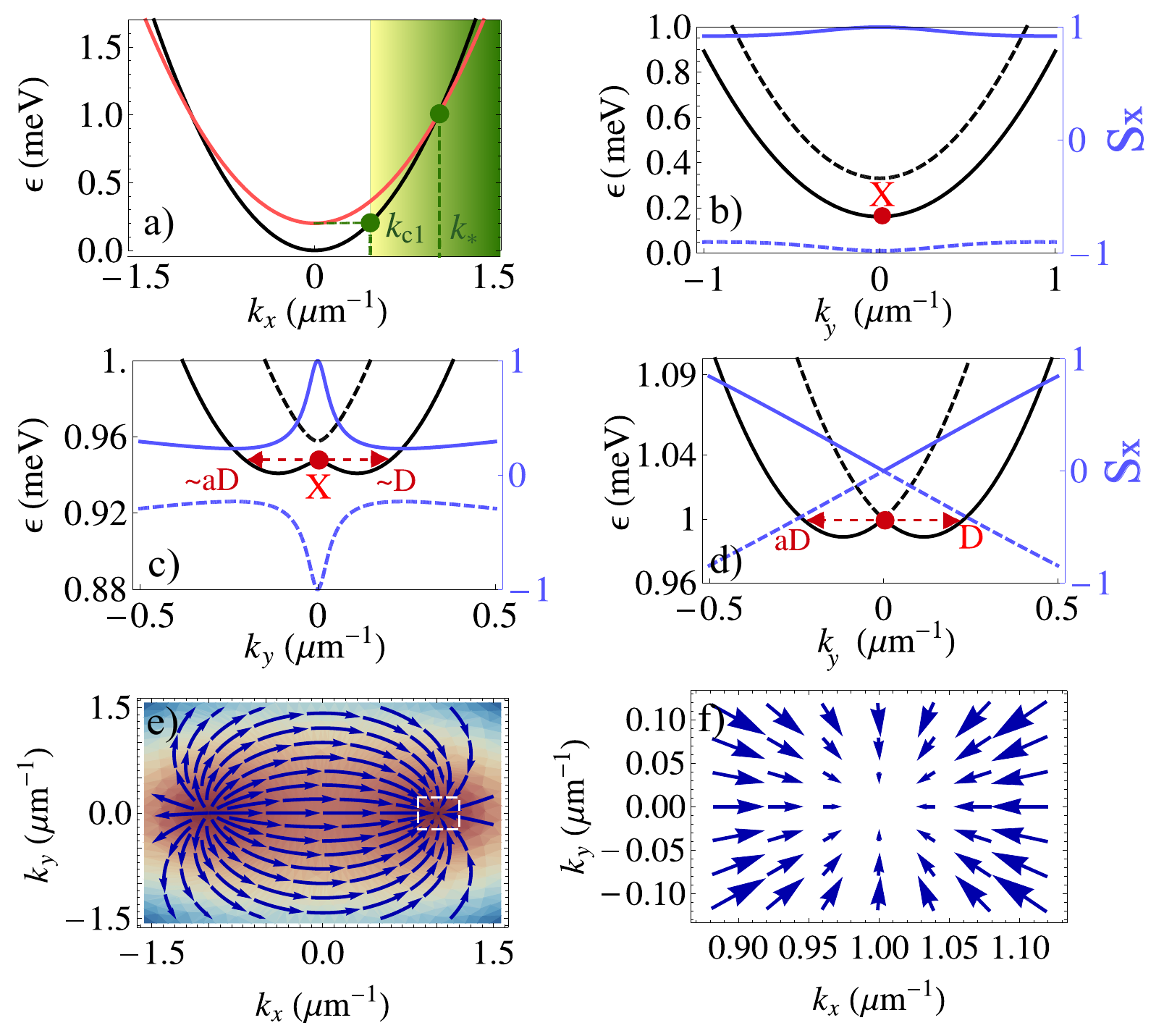}
\caption{(color online) Polariton dispersion relation for $\varphi=0$ (relative to $\Omega/2$). a) Dispersion along the $x$-direction showing the two polariton branches $\epsilon_{\pm}$ crossing at the magic point $k_*=\sqrt{\Omega/2\beta}$. The shadowed rectangle defines the region propagation $\vert \mathbf{k}_0-\mathbf{k}_*\vert \leq k_{c1}$, where the transverse dispersion bends. Panels b), c) and d) the black curves shows  the dispersion along the $y$-direction and the blue (gray) curves the corresponding x-projection of the pseudospin, respectively for $k_x=\left\{0.4,0.97, 1.0\right\}$ $\mu$m. In c), a parametric process $k_y:0\rightarrow \pm \kappa_p$ occurs between a X-polarized mode and a diagonal (D) and an anti-diagonal (aD) mode. In d), we observe the signature of the gauge field, with the two branches crossing each other at $k_y=0$. Panel e) depicts the field texture of (\ref{single}) in the reciprocal space and panel f) zooms around the wavevector $k_*$. We have considered the following parameters: $m=4.5 \times 10^{-5}m_e$, $\Omega=0.2$ meV and $\beta=0.1$ meV$\mu$m$^2$ (with $m_\ell\simeq 0.88m_t$), yielding $k_*= 1.0~\mu$m$^{-1}$.}
\label{fig2}
\end{figure}
\begin{figure}[t!]
\includegraphics[scale=0.15]{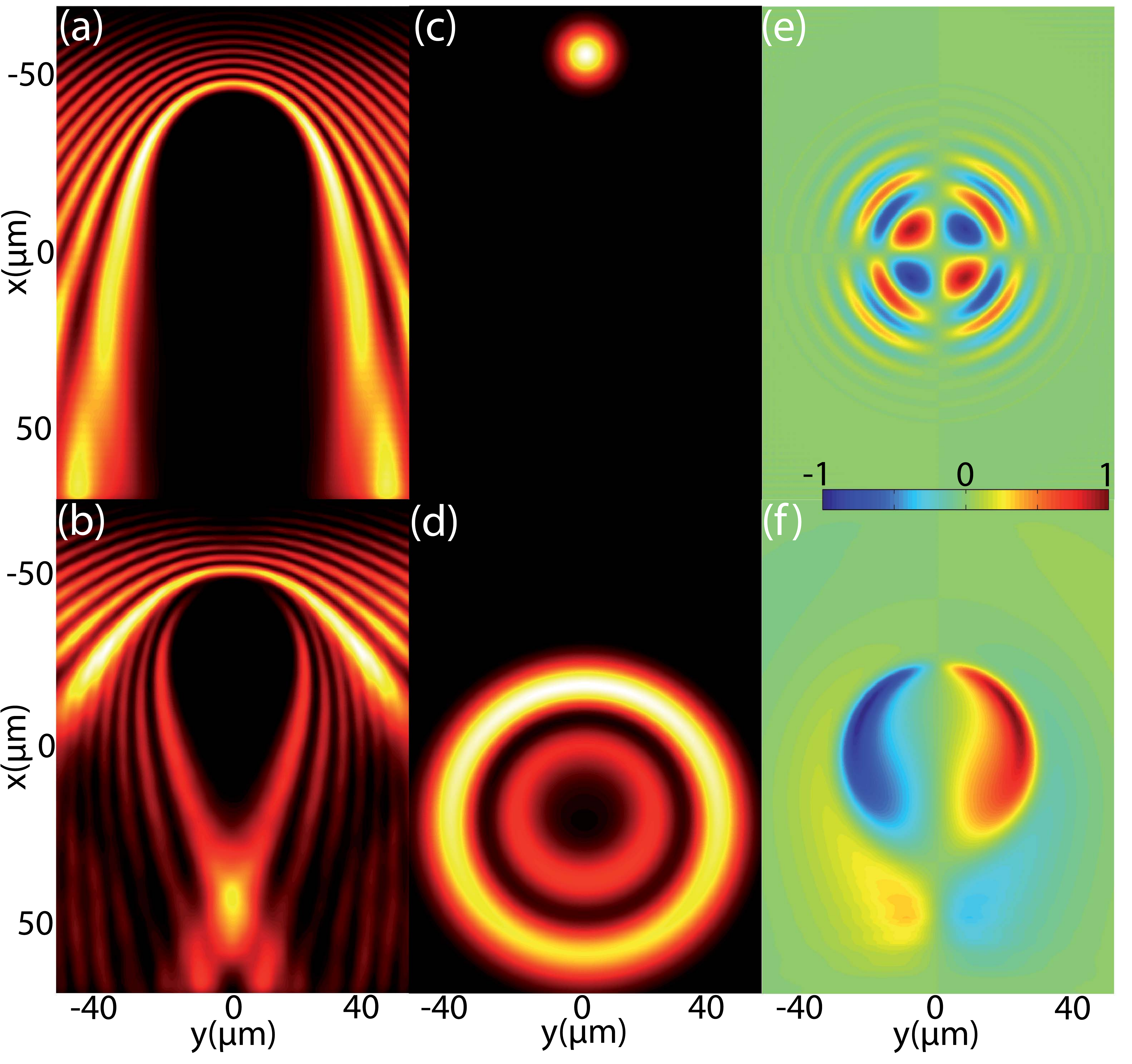}
\caption{(Color online) Resonant excitation of the magic point of the dispersion: (a)-(b) Lensing effect: Propagation at $\mathbf{k}=\mathbf{k}_*$ of the noninteracting linearly polarized polariton fluid against a Gaussian potential barrier (a) in the absence and (b) in the presence of the effective magnetic fields. The colormap shows the total density $n=n_++n_-$. (c)-(d) Conical diffraction under circularly polarized pulsed excitation at $\mathbf{k_0}=\mathbf{k}_*$. (c) $t=0$ ps and (d) $t=50$ ps. (e)-(f) Polarization domains. Gaussian linearly polarized pulsed excitation (e) at $\mathbf{k}_0=\mathbf{0}$ for $\Omega=0$ and (f) $\mathbf{k}=\mathbf{k}_*$ for $\Omega \neq 0$. The colormap displays here the degree of circular polarization $\rho_c=(n_++n_-)/(n_+-n_-)$.}
\label{fig3}
\end{figure}

We consider the motion of a bosonic fluid described by the spinor Gross-Pitaevskii equation
\begin{equation}
i\hbar\frac{\partial \Psi}{\partial t}=\hat H_0\Psi+\alpha_1\Psi^\dagger\Psi\Psi +\alpha_2(\sigma_x\Psi^\dagger\Psi)\Psi,
\label{GP1}
\end{equation}
where $\Psi=(\psi_+,\psi_-)^T$, $\alpha_1$ and $\alpha_2$ are the intra- and inter-spin interaction constants, and $\sigma_x$ is the anti-diagonal Pauli matrix. The single particle Hamiltonian, valid for both weakly coupled photons and polaritons in the parabolic approximation, is written as
\begin{equation}
\hat H_0=\left[
\begin{array}{cc}
\displaystyle{-\frac{\hbar^2\nabla^2}{2m}} & \displaystyle{-\frac{\Omega}{2} e^{-i\varphi} + \beta\left(\partial_y-i\partial_x\right)^2}\\\\
\displaystyle{-	\frac{\Omega}{2} e^{i\varphi}+ \beta\left(\partial_y+i\partial_x\right)^2} &\displaystyle{ -\frac{\hbar^2\nabla^2}{2m}}
\end{array}
\right],
\label{single}
\end{equation}
where $\beta=\hbar^2/(2m_r)$ is the strength of the TE-TM field, $m_\ell$ and $m_t$ represent the longitudinal and transverse polariton masses, $m_r=m_\ell m_t/(m_t-m_\ell)$ and $m=m_\ell m_t/(m_\ell+m_t)$ is the reduced polariton mass. The Hamiltonian (\ref{single}) describes an ideal polariton BEC, in the limit of the very long life-time, decoupled from the thermal bath. Diagonalization  leads to the following single-particle spectrum
\begin{equation}
\epsilon_\pm=\frac{\hbar^2 k^2}{2m}\pm \sqrt{\beta^2 k^4-\beta\Omega k^2\cos(2\theta_k-\varphi)+\frac{\Omega^2}{4}},
\label{sp1}
\end{equation}
where $k=\sqrt{k_x^2+k_y^2}$ and $\mathbf{k}=k(\cos\theta_k,\sin\theta_k)$. Due to the anisotropy of the TE-TM field, Eq.(\ref{sp1}) encodes very interesting features, as summarized in Fig. \ref{fig2}. 
Let us take $\varphi=0$, for a moment. Fig. \ref{fig2} a) depicts Eq. (3) along $k_x$ ($\theta_k=0)$, exhibiting the crossing of the two branches at $k_*=\sqrt{\Omega/2\beta}$, the point at which the TE-TM field and $\bm\Omega$ exactly compensate. For $k_x<k_{c1}$, with $k_{c1}=\sqrt{\Omega m_\ell/2\beta m_t}$, the energy is a monotonic function of $k_y$ (Fig. \ref{fig2}b)); on the other hand, for $k_x>k_{c1}$ the lower branch $\epsilon_-(k_y)$ bends and the parametric process $k_y:0\rightarrow  \pm\kappa_p$, with
\begin{equation}
\kappa_p=\left\{\frac{k_0^2/\left(1+m/2m_r\right)-k_*^2\left(1-m/2m_r\right)}{1-m/4m_r}\right\}^{1/2},
\end{equation}
conserving both energy and momentum is possible (see Fig. \ref{fig2} c)). As we will see later on, this resonance leads to a transverse instability of a superfluid flow. The dispersion is locally isotropic and linear near $q=0$, $\epsilon_\pm-\epsilon(k_*) \simeq \hbar^2q^2/(2m)\pm \sqrt{\beta\Omega} q$, with $\mathbf{q}=\mathbf{k}-\mathbf{k}_*$. This corresponds to dispersion of a system containing a spin-orbit coupling of the Rashba type.
For arbitrary values of $\varphi$, the compensation takes place at the propagation wavevector $\mathbf{k}_*=\pm \sqrt{\Omega/2\beta}\left(\cos\varphi/2,\sin\varphi/2\right)$. Therefore, for particles propagating at a generic wave-vector $\mathbf{k}_0$, the photon Hamiltonian can be written as
\begin{equation}
\hat H_0=\frac{\hbar^2}{2m}\left(\mathbf{q}^2+\kappa_x\bm\sigma\cdot\mathbf{q}+\kappa_y\left(\bm\sigma\times\mathbf{q}\right)\cdot e_\mathbf{z}\right)-\frac{\delta}{2}\sigma_x,
\label{ham1}
\end{equation}
where $\kappa_x=2m\sqrt{\beta\Omega}/\hbar^2\cos\varphi/2$,$\kappa_y=2m\sqrt{\beta\Omega}/\hbar^2\sin\varphi/2$ and $\delta=2\beta(k_*^2-k_0^2)$. The bending in the dispersion, and therefore the signature of spontaneous symmetry breaking, occurs for $\vert \mathbf{k}_0-\mathbf{k}_*\vert \leq k_{c1}$ (shadowed region of Fig. \ref{fig2} a)). In that region, the terms scaling linearly in $q$ dominate for $q \simeq 0 $. Contrary to the TE-TM SOC in Eq.(\ref{single}), Hamiltonian \ref{ham1} describes a true gauge-field, minimally coupled Hamiltonian, where $\kappa_x$ and $\kappa_y$ represent the magnitude of the monopolar- and  Rashba- field coupling constants. The gauge field texture $\mathbf{S}(\mathbf{k})$ for the particular case of propagation along a horizontal magnetic field ($\theta=\varphi=0$) is illustrated in panel e) - and zoomed around $\mathbf{k}_*$ in panel f) - of Fig. \ref{fig2}. A simple rotation of the sample results in more general field (or spin) textures, related by the transformation $\mathbf{S}'(\mathbf{k})=\mathcal{R}(\varphi) \mathbf{S}[\mathcal{R}(-\varphi/2)\mathbf{k}]$, where $\mathcal{R}(\varphi)$ is the $2\times 2$ rotation matrix (see also \cite{supp}). Hereinafter, we consider the case $\varphi=0$ for definiteness. To illustrate the effect of gauge field in Eq. (\ref{ham1}), we consider photons (polaritons) propagating at the magic wave-number $\mathbf{k}_0=k_* e_\mathbf{x}$, i.e. $\delta=0$. A cylindrical defect of size $a$, described by the potential $U(r)=U_0\exp(-r^2/a^2)$, scatters elastically the flow with maxima.
Most of the light shows a weak wave vector change of the order of $2\pi/a$. The particles with positive $q_y$ are pulled down, and those with negative $q_y$ are pushed up (see supplementary  \cite{supp} for more details). In fact, the defect focuses the beam, as if light was ``bent" by it (see Figs. \ref{fig3} a) and \ref{fig3} b)). This is quite remarkable, since focusing is obtained with an impenetrable defect. Another illustration presented on the Figs. \ref{fig3} c) and d) is the occurrence of conical diffraction.  The physical origin of Hamilton's conical diffraction \cite{Hamilton} is a singularity in $k$ space (diabolical point) due to polarization splitting (birefringence) \cite{Lloyd, Berry}, which can be obtained for our system with a circularly polarized pump. Dirac points in graphene \cite{Peleg} or the crossing point of a Rashba Hamiltonian are examples of diabolical points. A third signature of Eq.(\ref{ham1}), also being easily testable in basic experiments, is the conversion of a four-fold spin pattern, associated with spin Hall effect due to the TE-TM field \cite{Leyder, hugof} into a two-fold pattern (see Fig. \ref{fig3} e) and f)) induced by the fields symmetries.

\par
\begin{figure}[t!]
\includegraphics[scale=0.44]{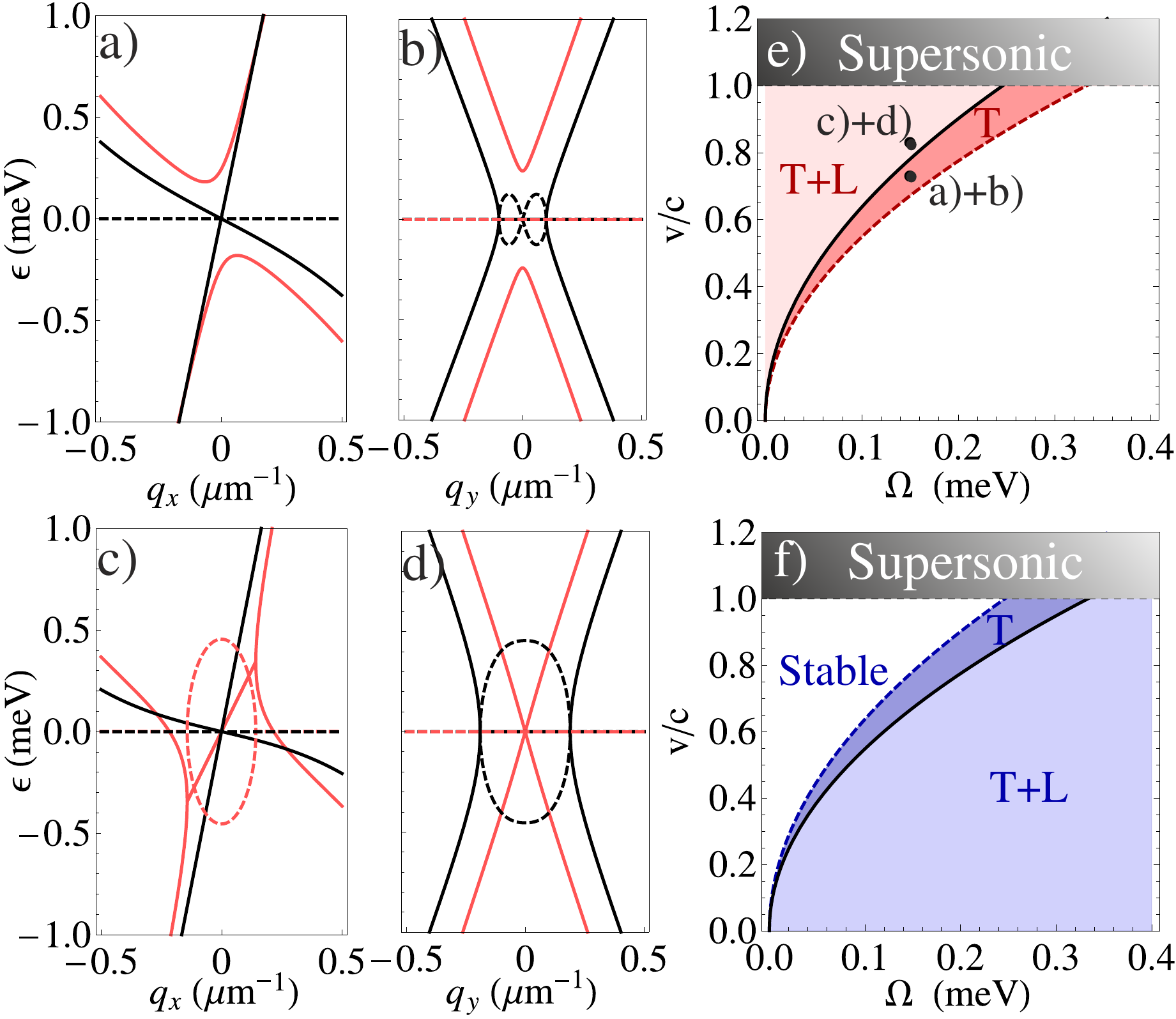}
\caption{(color online) From a) to d): Transverse (T) and longitudinal (L) instabilities for a X- polarized condensate in the subsonic regime with a blue-shift of $\alpha_1n=2.5$ meV and $\alpha_2=-0.2\alpha_1$. The solid and dashed lines respectively represent the real and imaginary parts. Black (gray) lines depicting the lower (upper) branches $\epsilon_{x,y}^L$ ($\epsilon_{x,y}^U$).  Transverse (T) instability: $\epsilon_x$ (panel a)) and $\epsilon_y$ (panel b)) for $k_0=0.93k_*>k_{c1}$.  Transverse+Longitudinal (T+L) instability: $\epsilon_x$ (panel c)) and $\epsilon_y$ (panel d)) for $k_0=1.12k_*$. Stability diagram for X- (panel e)) and Y-polarized (panel f)) condensates. The black solid line represents the velocity $v_*=\hbar k_*/m$, and the red/light grey (blue/light grey) dashed lines correspond to $v_{c1}=\hbar k_{c1}/m$ ($v_{c2}=\hbar k_{c2}/m$). The points correspond to the parameters used in panels a)/ b) and c)/d). We have considered $\beta=0.15$ meV$\mu$m$^2$.}
\label{fig4}
\end{figure}

 
In what follows, we investigate the effect of inter-particle interactions. First, in contrast to the situation in atomic BECs, with linear in $k$ SOC \cite{wang}, an homogeneous, linearly polarized polariton condensate at rest remains stable in the presence of the SOC introduced in Eq. (\ref{single}) for $\Omega>0$, provided that $m<m_r$ and $\vert\alpha_1\vert>\vert\alpha_2\vert$, which is the case in experimental situations. The particular case $\Omega=0$ has been investigated in \cite{shelykh}. The situation is dramatically modified if the homogeneous condensate is put in motion in the vicinity of the ``magic" point $k_*$. The collective excitations of a polariton BEC propagating with velocity $\mathbf{v}_0=\hbar \mathbf{k}_0/m$ can be determined by linearizing (\ref{GP1}) with  $\bm \Psi=e^{i(\mathbf{k}_0\cdot \mathbf{r}-\mu t/\hbar)}\left[\bm \Psi_0+\sum_{\mathbf{q}}(\mathbf{u}_\mathbf{q}e^{i(\mathbf{q}\cdot \mathbf{r}-\omega t)}+\mathbf{v}^*_\mathbf{q} e^{-i(\mathbf{q}\cdot \mathbf{r}-\omega t)})\right]$ \cite{supp}, for a linearly polarized condensate, $\bm\Psi_0=(n_+,n_-)^T=n_0(e^{i\eta},e^{-i\eta})^T$, with $\eta$ being the polarization angle. A condensate is X- (Y-) polarized for $\eta=0$ ($\eta=\pi/2$), and the associated pseudo-spin is $\mathbf{S}_0=n_0 e_\mathbf{x}$ ($\mathbf{S}_0=-n_0 e_\mathbf{x}$) \cite{flayac}. The chemical potential contains the interaction and ``magnetic" energy of the condensate
\begin{equation}
\mu=(\alpha_1+\alpha_2)n_0-\frac{1}{2}\left(\Omega-2 \beta k_0^2\cos2\theta\right)\cos2\eta.
\end{equation}
We consider a BEC propagating with wavevector $\mathbf{k}_0=k_0e_\mathbf{x}$ and for which the non-linearity is large enough so the critical Landau velocity for the breakdown of superfluidity is larger than $v_0=\hbar k_0/m$. Let us discuss the case of a X-polarized BEC first (X is the ground state at $k_x=0$).  The spectrum of the elementary excitations contains four anisotropic branches - two with positive and two negative energies -, $\pm\epsilon^{L,U}(\mathbf{q})$, where the Doppler-shifted lower (L) and upper (U) branches can be written in the long wavelength limit $q\sim0$ as 
\begin{equation}
\begin{array}{c}
\displaystyle{\epsilon^L_x\simeq  \hbar c_\ell^L \vert q_x\vert +\frac{\hbar^2 k_0}{m_\ell}q_x, \quad 
\epsilon^L_y \simeq \hbar c_\ell^L \sqrt{\frac{k_{c1}^2-k_0^2}{k_{*}^2-k_0^2}}\vert q_y \vert }\\
\displaystyle{ \epsilon^U_x \simeq \left(\Delta_1+\frac{\hbar^2q_x^2}{2 m_t}\zeta\right)+\frac{\hbar k_0}{m_t}q_x, \quad \epsilon^U_y\simeq  \left\{ \Delta_1^2+\frac{\hbar^2 q_y^2}{m_t}\Delta_2\right\}^{1/2}},
\end{array}
\label{modesBEC}
\end{equation}
where $c_{\ell,t}^{L}=[(\alpha_1+\alpha_2)n/m_{\ell,t}]^{1/2}$ are the two speeds,  $\Delta_1=\sqrt{\delta[\delta +2(\alpha_1-\alpha_2)n]}$, $\zeta=[\delta +(\alpha_1-\alpha_2)n]/\Delta_1$, $\delta=2\beta(k_*^2-k_0^2)$ and $\Delta_2$ can be found in Ref. \cite{supp}. Eq. (\ref{modesBEC}) allows us to understand the onset of both transverse and longitudinal instabilities, as illustrated in Fig. \ref{fig4}. 
For $k<k_{c1}$, all the branches are real and no instability takes place. The X-polarized condensate is therefore superfluid. For $k_{c1}<k_0<k_*$, $\epsilon^L_y$ becomes complex,  corresponding to the onset of a transverse-instability (T), as depicted in Fig. \ref{fig4} a) and b). The most unstable mode corresponds to the wave-vectors $q_y=\pm\kappa_p/2$, where $\kappa_p$ is the mode for which the single-particle spectrum admits the parametric process conserving both energy and momentum (see Fig. \ref{fig1}). For $k_0>k_*$, $\delta <0$ and $\Delta_1$ is imaginary, which implies an instability in the upper branch $\epsilon^U_x$. This corresponds to the onset of a longitudinal instability (L), as plotted in Fig. \ref{fig4} c). We notice that the latter is accompanied by a transverse instability  (T+L-instability), as can be seen in Fig. \ref{fig4} d). The most unstable mode, however, is now observed for $(q_x, q_y)=(0,\kappa_p/2)$, similar to the zero wave-vector  instability  discussed for atomic BECs (see e.g. Ref.\cite{wang}). The substantial difference is that here this instability is only achieved for a propagating (and not for a static) condensate. The phase diagram summarizing these results is depicted in Fig. \ref{fig4} e). Increasing the velocity, the system evolves from superfluid to transversally, transversally and longitudinally unstable to finally reach the supersonic regime $v_0>c$, with $c=c_\ell c_t/(c_\ell+c_t)$.\par

The even more intriguing case of a Y-polarized BEC is summarized in Fig. \ref{fig4} f). For very low wave-vectors, namely $k_0<k_{c1}$, the spin is anti-aligned with the static field ($\mathbf{S}=-n_0 e_\mathbf{x}$) and the system is T+L- unstable. Above $k_*$, the imaginary part of the energy vanishes in the $x$ direction, and the system exhibits a purely transverse instability. At wave-vectors higher than the second critical wave-vector $k_{2c}=k_*\sqrt{m_t/m_\ell}$, the system becomes stable. This counter-intuitive result simply states that superfluidity breaks down when the flow velocity is decreased or when particle density is increased.

\begin{figure}[t!]
\centering
\includegraphics[scale=0.4]{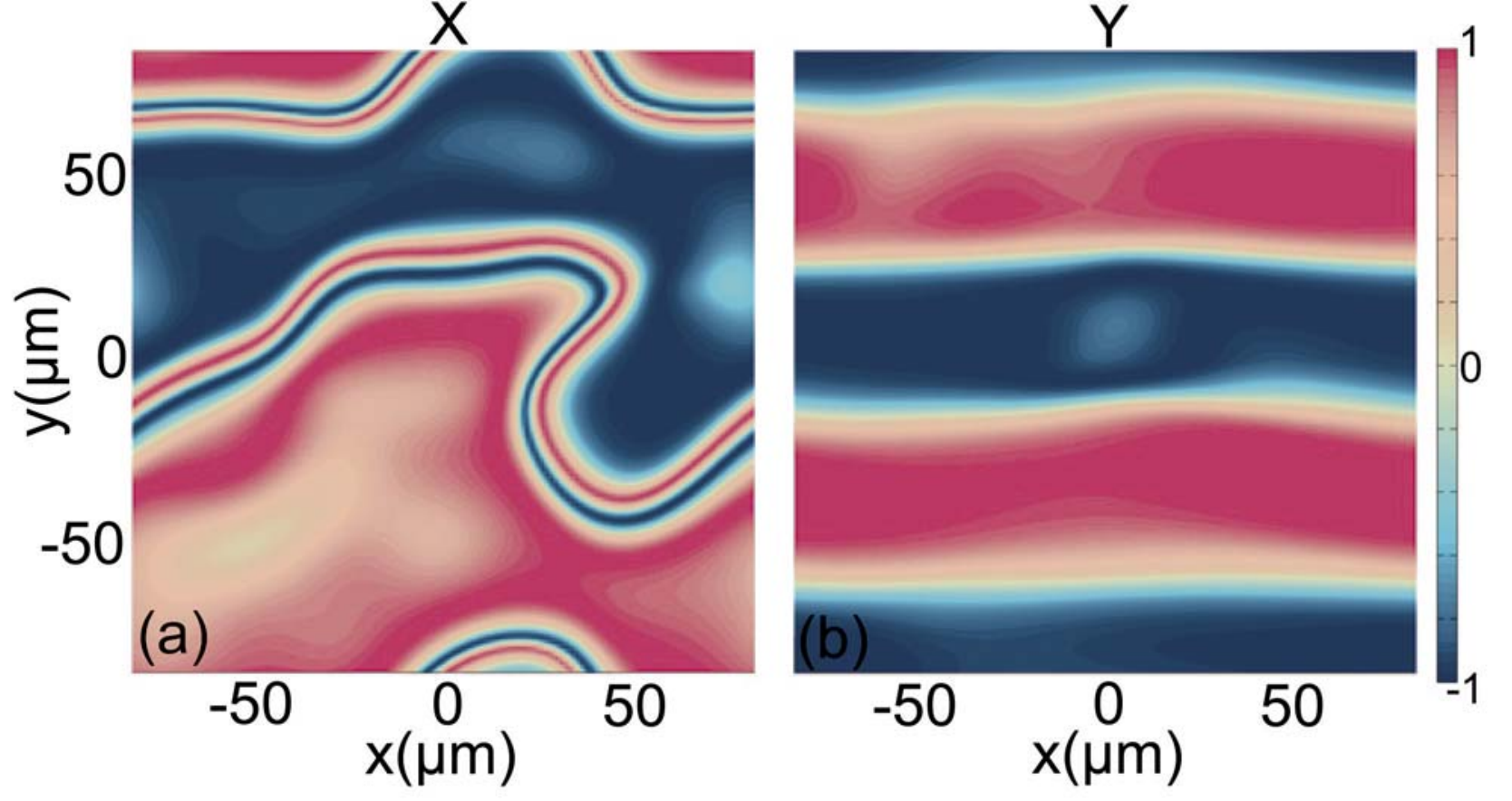}
\caption{ (color online) Snapshots of the diagonal polarization degree $\rho_D=(n_D-n_{aD})/(n_D+n_{aD})$ after the onset of instabilities in X-polarized BEC. $v_0/c=0.45$ (a) and $0.6$ (b).}
\label{fig5}
\end{figure}

To confirm the previous results, we performed numerical simulations using Eq. (\ref{GP1}). For X-polarized condensates, we observe the formation of stripes along the transverse direction for $k_0>k_{c1}$, illustrating the onset of a transverse instability (Fig. \ref{fig5} a)) and the formation of spin domain walls for $k_0>k_*$ through the T+L- instability mechanism (see Fig. \ref{fig5} b)). For Y-polarized BEC, the former is observed for $k_*>k_0>k_{c2}$ while the latter is obtained for $k_0<k_*$ (see \cite{supp}). At late stages, the instability saturates and the domain wall breaks into pair of half-vortices, the elementary topological defects in spin-anisotropic BECs \cite{Rubo} (a movie is provided in \cite{supp}). For both polarizations, we observe that the formation of spin domain walls is much faster than the formation of stripes, which can be understood from the polarization of the excitations. The onset of  instability is the consequence of a parametric process which populates the modes $ q_y=\pm \kappa_p$ ($ q_x=\pm \kappa_x$) from the initial state at $(q_x, q_y)=(0,0)$. The efficiency of the process is proportional to $\cos^2(2\Delta\eta)$, where $\Delta\eta$ is the relative polarization angle between the initial and final states \cite{opo, ciuti}. From Fig. \ref{fig2}, we observe that a process of the type $q_y:0\rightarrow \pm \kappa_p$ occurs between an X- and two almost diagonal ($\sim$D and $\sim$aD) states, for which $\Delta\eta\simeq \pi/4$ and $\cos^2(2\Delta\eta)\ll1$. The longitudinal instability develops faster, since the process $q_x:0 \rightarrow \pm\kappa_x$ conserves polarization and, therefore, $\cos^2(2\Delta\eta)=1$. Our results reveal a curious difference with respect to atomic BECs, where exotic phases are formed in static systems \cite{wang, santos, hu}. 
\par
To conclude, we have shown that a non-abelian effective gauge field appears for a propagating polariton condensate as an interplay of the TE-TM splitting and a constant in-plane field. The configuration we propose is realistic and can be easily realized in microcavities where both static and TE-TM fields are present. This gauge field leads to lensing effects, conical diffraction, polarization patterns and to the instability of superfluid flow accompanied by the formation of spin-textured states.

We thank I. A. Shelykh for discussions. This work was supported by ``ANR QUANDYDE", the Labex Gannex, FP7 ITN ``Spin-Optronics"(237252).


\begin{thebibliography}{99}

\bibitem{lin1}
Y.-J. Lin, R. L. Compton, K. Jim\'enez-Garc\'ia, J. V. Porto, and I. B. Spielman, Nature {\bf 462}, 628 (2009).

\bibitem{lin2}
Y.-J. Lin,  K. Jim\'enez-Garc\'ia, and I. B. Spielman, Nature {\bf 471}, 83 (2011).

\bibitem{aidelsburger}
M. Aidelsburger, M. Atala, S. Nascimb\`ene, S. Trotzky, Y.-A. Chen, and I. Bloch, Phys. Rev. Lett. {\bf 107}, 255301 (2011).

\bibitem{chen}
S. Chen, J.-Y. Zhang, S.-C. Ji, Z. Chen, L. Zhang, Z.-D. Du, Y. Deng, H. Zhai, and J.-W. Pan, arXiv:1201.6018.

\bibitem{hall1}
Y. K. Kato, R. C. Meyers, A. C. gossard, and D. D. Awschalom, Science {\bf 306}, 1910 (2004).

\bibitem{hall2}
M. Konig et al., Science {\bf 318}, 766 (2007).

\bibitem{ti1}
C. L. Kane, and E. J. Mele, Phys. Rev. Lett. {\bf 95}, 146802 (2005).

\bibitem{ti2}
B. A. Bernevig, T. L. Hughes, and S.-C. Zhang, Science {\bf 314}, 1757 (2006).

\bibitem{koralek}
J. D. Koralek et al., Nature {\bf 458}, 610 (2009).

\bibitem{Carusotto}
R.O. Umucalilar, I. Carusotto, Phys. Rev. A {\bf 84}, 043804 (2011).

\bibitem{Hafezi}
M. Hafezi, E. A. Demler, M. P. Lukin, and J. M. Taylor, Nat. Phys. {\bf 7}, 907 (2011).

\bibitem{Hafeziarxi}
M. Hafezi, J. Fan, A. Migdall, J. Taylor, arXiv:1302.2153 (2013).

\bibitem{Fang}
K. Fang, Z. Yu, and S. Fan, Opt. Exp. {\bf 21}, 18216 (2013).

\bibitem{Rechtman}
M. C. Rechtsman, J. M. Zeuner, A. Tnnermann, S. Nolte, M. Segev, and A. Szameit, Nat. Photon. {\bf 7}, 153 (2013).

\bibitem{Rechtmannature}
 M. C. Rechtsman, J. M. Zeuner, Y. Plotnik,	Y. Lumer, D. Podolsky, F. Dreisow, S. Nolte, M. Segev and A. Szameit, Nature {\bf 496}, 196 (2013).
 
\bibitem{Jia}
N. Jia, A. Sommer, D. Schuster, J. Simon, arXiv:1309.0878 (2013).

\bibitem{review}
I.A. Shelykh, A.V. Kavokin, Y.G. Rubo, T.C.H. Liew, G. Malpuech, Semic. Sc.\& Techn. {\bf 25}, 013001, (2010).

\bibitem{kuther}
A. Kuther, M. Bayer, T. Gutbrod, A. Forchel, P. A. Knipp, T. L. Reinecke, and R. Werner, Phys. Rev.B {\bf 58}, 15744 (1998).

\bibitem{Klopo}
L. Klopotowski, M. D. Martin, A. Amo, L. Vina, I. A. Shelykh, M. M. Glazov, G. Malpuech, A. V. Kavokin, and R. Andre, Solid State Com., {\bf 139}, 511 (2006).


\bibitem{hugof}
H. Flayac, D. D. Solnyshkov, I. A. Shelykh, and G. Malpuech, Phys. Rev. Lett. {\bf 110}, 016404 (2013).

\bibitem{OSHE}
A. Kavokin, G. Malpuech, and M. Glazov, Phys. Rev. Lett. {\bf 95}, 136601 (2005).

\bibitem{Leyder}
C. Leyder, M. Romanelli, J. Ph. Karr, E. Giacobino, T. C. H. Liew, M. M. Glazov, A. V. Kavokin, G. Malpuech, A. Bramati, Nature Physics {\bf 3}, 628 (2007).

\bibitem{Soln}
D. D. Solnyshkov, H. Flayac, and G. Malpuech, Phys. Rev. B {\bf 85}, 073105 (2012).

\bibitem{Hivet}
R. Hivet, H. Flayac, D.D. Solnyshkov, D. Tanese, T. Boulier, D. Andreoli, E. Giacobino, J. Bloch, A. Bramati, G. Malpuech, A. Amo, Nature Physics  {\bf 8}, 724 (2012).


\bibitem{shelykh}
I. A. Shelykh, Y. G. Rubo, G. Malpuech, D. D. Solnyshkov, and A. Kavokin, Phys. Rev. Lett. {\bf 97}, 066402 (2006).


\bibitem{dalibard}
J. Dalibard, F. Gerbier, G. Juzeli\"unas, P. \"Ohberg, Rev. Mod. Phys. {\bf 83}, 1523 (2011).

\bibitem{rashba}
Y. A. Bychkov and E. I. Rashba, J. Phys. C {\bf 17}, 6039 (1984).

\bibitem{dresselhaus}
G. Dresselhaus, Phys. Rev. {\bf 100}, 580 (1955).

\bibitem{wang} C. Wang, C. Gao, C.-M. Jian, and H. Zhai, Phys. Rev. Lett. 105, 160403 (2010).

\bibitem{santos}
S. Sinha, R. Nath, and L. Santos, Phys. Rev. Lett. {\bf 107}, 270401 (2011).

\bibitem{hu}
H. Hu, B. Ramachandhran, H. Pu, and X-Ji Liu, Phys. Rev. Lett. {\bf 108}, 010402 (2012).


\bibitem{Hamilton}
 W. R. Hamilton, Trans. R. Irish Acad. {\bf 17}, 1 (1837).

\bibitem{Lloyd}
H. Lloyd, Trans. R. Irish Acad. {\bf 17}, 145 (1837).

\bibitem{Berry}
M.V. Berry, M.R. Jeffrey and J.G. Lunney, Proc. R. Soc. A {\bf 462}, 1629 (2006).

\bibitem{Peleg}
Or Peleg et al. Phys. Rev. Lett. {\bf 98}, 103901 (2007).

\bibitem{supp}
See Suplemental Material for a schematic explanation of the lensing effect and further details on the linearization of Eq.(\ref{GP1}).

\bibitem{flayac}
H. Flayac, D. D. Solnyshkov and G. Malpuech, Phys. Rev. B {\bf 83}, 193305 (2011).

\bibitem{Rubo} Yu.G. Rubo, Phys. Rev. Lett. 99, 106401 (2007).


\bibitem{opo}
A. Kavokin and G. Malpuech, 5th chapter of Cavity Polaritons, Edited by V.M. Agranovich, Elsevier (2003).

\bibitem{ciuti}
I. Carusotto, and C. Ciuti, Rev. Mod. Phys. {\bf 85}, 299 (2013).





\end{thebibliography}
\end{document}